\documentclass[11pt]{elsart}
\usepackage{color}
\usepackage{epsfig}
\usepackage{amssymb}

\newcommand\fverb{\setbox\pippobox=\hbox\bgroup\verb}
\newcommand\fverbdo{\egroup\medskip\noindent%
            \fbox{\unhbox\pippobox}\ }
\newcommand\fverbit{\egroup\item[\fbox{\unhbox\pippobox}]}
\newbox\pippobox
\newcommand{\eeg}{$e^+ e^- \gamma$}
\newcommand{\mmg}{$\mu^+ \mu^- \gamma$}

\newcommand{\ppp}{$\pi^+ \pi^- \pi^0$}
\def\ifm#1{\relax\ifmmode#1\else$#1$\fi}
\def\ff{$\phi$--factory}  \def\DAF{DA\char8NE}
\def\up#1{$^{#1}$}  \def\dn#1{$_{#1}$}  \def\dif{\hbox{d\kern.1mm}}
\def\po{\ifm{\pi^0}}    \def\f{\ifm{\phi}}
\def\pic{\ifm{\pi^+\pi^-}} \def\epm{\ifm{e^+e^-}}  \def\to{\ifm{\rightarrow\,}}
\def\plm{\ifm{\pm}}    \def\deg{\ifm{^\circ}}  
\def\gam{\ifm{\gamma}}  \def\x{\ifm{\times}}  \def\ab{\ifm{\sim}}  \def\sig{\ifm{\sigma}}
\def\ie{{\it\kern-1pt i.\kern-.5pt e.\kern-.2pt}}  
\let\cl=\centerline  \def\L{\ifm{{\cal L}}}  \def\Dam{\ifm{\Delta^{\pi\pi} a_\mu}}
\def\lfigbox#1;#2;{\parbox{#2cm}{\vglue3mm\epsfig{file=#1.eps,width=#2cm}\vglue3mm}}
\def\km{\kern-1.5mm}  \def\kak{\km&\km}  \def\kma{\kern-2.5mm}\def\kms{\kern-.75mm}
\def\pt#1,#2,{\ifm{#1\x10^{#2}}}   \def\bye{\end{document}}
\def\strike#1;{\setbox0=\hbox{#1}\rlap{\vrule height3.3pt depth-2.3pt width\wd0}\kern2pt#1\ }

\catcode`@=11 
\newdimen\z@ \z@=0pt 
\newskip\z@skip \z@skip=0pt plus0pt minus0pt
\def\m@th{\mathsurround=\z@}
\def\ialign{\everycr{}\tabskip\z@skip\halign} 
\def\eqalign#1{\null\,\vcenter{\openup\jot\m@th
  \ialign{\strut\hfil$\displaystyle{##}$&$\displaystyle{{}##}$\hfil
      \crcr#1\crcr}}\,}
\catcode`@=12 
\newcommand{\eV}{{e\kern-.07em V}}

\newcommand{\MeV}{{\rm \,M\eV}}

\newcommand{\GeV}{{\rm \,G\eV}}

\newcommand{\dd}{\ensuremath{{\rm d}}}

\definecolor{dmagen}{rgb}{.7,.0,.7}
\definecolor{dgree}{rgb}{.0,.5,.0}
\definecolor{dred}{rgb}{.9,.0,.0}

\makeatletter
\def\@ptsize{1}
\belowdisplayskip \abovedisplayskip
\abovedisplayshortskip \z@ \@plus 2\p@%
\def\section{\@startsection{section}{1}{.5\z@}{1\@bls
  \@plus .1\@bls \@minus .1\@bls}{.5\@bls}{\normalsize\bfseries}}
\def\subsection{\@startsection{subsection}{2}{.5\z@}{.3\@bls
  \@plus .1\@bls \@minus .1\@bls}{.3\@bls}{\normalsize\itshape}}
\makeatother

\begin{document}

\begin{frontmatter}

\title{\mathversion{bold} Measurement of $\sigma(e^+ e^- \rightarrow\pi^+\pi^-\gamma(\gamma))$ and the dipion
contribution to the muon anomaly with the KLOE detector}
\collab{The KLOE Collaboration}
\author[Na,infnNa]{F.~Ambrosino},
\author[Frascati]{A.~Antonelli},
\author[Frascati]{M.~Antonelli},
\author[Roma2,infnRoma2]{F.~Archilli},
\author[Roma3,infnRoma3]{C.~Bacci},
\author[Karlsruhe]{P.~Beltrame},
\author[Frascati]{G.~Bencivenni},
\author[Frascati]{S.~Bertolucci},
\author[Roma1,infnRoma1]{C.~Bini},
\author[Frascati]{C.~Bloise},
\author[Roma3,infnRoma3]{S.~Bocchetta},
\author[Frascati]{F.~Bossi},
\author[infnRoma3]{P.~Branchini},
\author[Frascati]{P.~Campana},
\author[Frascati]{G.~Capon},
\author[Frascati]{T.~Capussela},
\author[Roma3,infnRoma3]{F.~Ceradini},
\author[Frascati]{S.~Chi},
\author[Na,infnNa]{G.~Chiefari},
\author[Frascati]{P.~Ciambrone},
\author[Roma1]{F.~Crucianelli},
\author[Frascati]{E.~De~Lucia},
\author[Roma1,infnRoma1]{A.~De~Santis},
\author[Frascati]{P.~De~Simone},
\author[Roma1,infnRoma1]{G.~De~Zorzi},
\author[Karlsruhe]{A.~Denig},
\author[Roma1,infnRoma1]{A.~Di~Domenico},
\author[infnNa]{C.~Di~Donato},
\author[Pisa]{S.~Di~Falco},
\author[Roma3,infnRoma3]{B.~Di~Micco},
\author[infnNa]{A.~Doria},
\author[Frascati]{M.~Dreucci},
\author[Frascati]{G.~Felici},
\author[Frascati]{A.~Ferrari},
\author[Frascati]{M.~L.~Ferrer},
\author[Roma1,infnRoma1]{S.~Fiore},
\author[Frascati]{C.~Forti},
\author[Roma1,infnRoma1]{P.~Franzini},
\author[Frascati]{C.~Gatti},
\author[Roma1,infnRoma1]{P.~Gauzzi},
\author[Frascati]{S.~Giovannella},
\author[Lecce,infnLecce]{E.~Gorini},
\author[infnRoma3]{E.~Graziani},
\author[Pisa]{M.~Incagli},
\author[Karlsruhe]{W.~Kluge},
\author[Moscow]{V.~Kulikov},
\author[Roma1,infnRoma1]{F.~Lacava},
\author[Frascati]{G.~Lanfranchi},
\author[Frascati,StonyBrook]{J.~Lee-Franzini},
\author[Karlsruhe]{D.~Leone},
\author[Frascati,Moscow]{M.~Martemianov},
\author[Frascati,Energ]{M.~Martini},
\author[Na,infnNa]{P.~Massarotti},
\author[Frascati]{W.~Mei},
\author[Na,infnNa]{S.~Meola},
\author[Frascati]{S.~Miscetti},
\author[Frascati]{M.~Moulson},
\author[Frascati]{S.~M\"uller\corauthref{cor}}\ead{stefan.mueller@lnf.infn.it},
\author[Frascati]{F.~Murtas},
\author[Na,infnNa]{M.~Napolitano},
\author[Roma3,infnRoma3]{F.~Nguyen\corauthref{cor}}\ead{nguyen@fis.uniroma3.it},
\author[Frascati]{M.~Palutan},
\author[infnRoma1]{E.~Pasqualucci},
\author[infnRoma3]{A.~Passeri},
\author[Frascati,Energ]{V.~Patera},
\author[Na,infnNa]{F.~Perfetto},
\author[infnLecce]{M.~Primavera},
\author[Frascati]{P.~Santangelo},
\author[Na,infnNa]{G.~Saracino},
\author[Frascati]{B.~Sciascia},
\author[Frascati,Energ]{A.~Sciubba},
\author[Frascati]{A.~Sibidanov},
\author[Frascati]{T.~Spadaro},
\author[Roma1,infnRoma1]{M.~Testa},
\author[infnRoma3]{L.~Tortora},
\author[infnRoma1]{P.~Valente},
\author[Karlsruhe]{B.~Valeriani},
\author[Frascati]{G.~Venanzoni\corauthref{cor}}\ead{graziano.venanzoni@lnf.infn.it},
\author[Frascati,Energ]{R.Versaci},
\author[Frascati,Beijing]{G.~Xu}
\address[Frascati]{Laboratori Nazionali di Frascati dell'INFN, Via E. Fermi 40, I-00044 Frascati, Italy.}
\address[Karlsruhe]{Institut f\"ur Experimentelle Kernphysik, Universit\"at Karlsruhe, D-76128 Karlsruhe,Germany.}
\address[Lecce]{Dipartimento di Fisica dell'Universit\`a del Salento, Via Arnesano, I-73100 Lecce, Italy.}
\address[infnLecce]{INFN Sezione di Lecce, Via Arnesano, I-73100 Lecce, Italy.}
\address[Na]{Dipartimento di Scienze Fisiche dell'Universit\`a  di Napoli ``Federico II'', Via Cintia, I-80126 Napoli, Italy.}
\address[infnNa]{INFN Sezione di Napoli, Via Cintia, I-80126 Napoli, Italy.}
\address[Pisa]{Dipartimento di Fisica dell'Universit\`a di Pisa, Largo Bruno Pontecorvo 3, I-56127 Pisa, Italy.}
\address[Energ]{Dipartimento di Energetica dell'Universit\`a di Roma ``La Sapienza'', P. Aldo Moro 2, I-00185 Roma, Italy.}
\address[Roma1]{Dipartimento di Fisica dell'Universit\`a di Roma ``La Sapienza'', P. Aldo Moro 2, I-00185 Roma, Italy.}
\address[infnRoma1]{INFN Sezione di Roma, P. Aldo Moro 2, I-00185 Roma, Italy.}
\address[Roma2]{Dipartimento di Fisica dell'Universit\`a di Roma ``Tor Vergata'',
Via della Ricerca Scientifica 1, I-00133 Roma, Italy.}
\address[infnRoma2]{INFN Sezione di Roma Tor Vergata, Via della Ricerca Scientifica 1, I-00133 Roma, Italy.}
\address[Roma3]{Dipartimento di Fisica dell'Universit\`a di Roma ``Roma Tre'', Via della Vasca Navale 84, I-00146 Roma, Italy.}
\address[infnRoma3]{INFN Sezione di Roma Tre, Via della Vasca Navale 84, I-00146 Roma, Italy.}
\address[StonyBrook]{Physics Department, State University of New York at Stony Brook, Stony Brook, NY 11794-3840  USA.}
\address[Beijing]{Institute of High Energy Physics of Academia Sinica,  P.O. Box 918 Beijing 100049, P.R. China.}
\address[Moscow]{Institute for Theoretical and Experimental Physics, B. Cheremushkinskaya ul. 25 RU-117218 Moscow, Russia.}
\begin{abstract}
We have measured the cross section $\sigma(e^+e^-\to\pic\gam(\gamma))$ at \DAF,\ the Frascati \ff, using events with
initial state radiation photons emitted at small angle and inclusive of final state radiation. 
We present the analysis of a new data set corresponding to an integrated luminosity of 240 pb$^{-1}$.
We have achieved a reduced systematic uncertainty
with respect to previously published KLOE results. From the 
cross section we obtain the pion form factor and the contribution to the muon magnetic anomaly from two-pion states in the mass range $0.592<M_{\pi\pi}<0.975$ GeV. For the latter we find $\Delta^{\pi\pi} a_\mu$=(387.2\plm0.5\dn{\rm stat}\plm2.4\dn{\rm exp}\plm2.3\dn{\rm th})\x10\up{-10}.
\end{abstract}\vglue-5mm
\vglue-5mm
\corauth[cor]{\noindent Corresponding Authors.}
\end{frontmatter}
\overfullrule6pt
\section{Introduction}
\label{sec:1}
The muon magnetic anomaly, $a_\mu$, has been recently measured at Brookhaven with an accuracy of 0.54 ppm \cite{Bennett:2006fi}.
The value of $a_\mu$ in the standard model, is found to differ from the experimental value by 2.8 to 3.4
standard deviations \cite{miller,Jegerlehner:2007xe}. The main source of uncertainty in the estimate of $a_\mu$ is the hadronic contribution,
which is not calculable in perturbative QCD.
The hadronic contribution, at lowest order, $\Delta^{\rm h,\,lo}a_\mu$, is obtained from a dispersive integral
over the cross section for \epm\to hadrons \cite{mich,deraf}.
The \epm\to\pic\ channel accounts for $\sim70\%$ of $\Delta^{\rm h,\,lo}a_\mu$ and $\sim60\%$ of its uncertainty.

It should be noted that the physically measurable cross section for \epm\to\pic\ , as such,
cannot be used in the dispersive integral for two reasons.
The first, obviously, is that the measured cross section is affected by initial state radiation (ISR) which must not be included in the
contribution to the muon anomaly. Even the energy at the $\pi\pi\gam$ vertex is different from the nominal \epm\ collision energy.
The second reason is more a question of tradition and book keeping. The photon at the \pic\gam\ vertex, at lowest order, is a bare photon,
\ie\ without vacuum polarization. The measured cross section must be therefore corrected for both effects, as we discuss later. 
Final state radiation (FSR) from the pions must instead be included. The measured quantities therefore require corrections for the
photon vacuum polarization, for ISR, and to ensure that pion FSR is included, since some of the events with FSR might have been rejected.
In our measurement there are some additional corrections, mostly due to ambiguities between ISR and FSR because
we measure the dipion mass and not the \epm\ collision energy.

In 2005, 
we published \cite{Aloisio:2004bu} a measurement of the dipion contribution \Dam, using the method described in
Sec. \ref{subsec:1.2}, using data collected in 2001 for $\int\!\L\dif t$=140 pb$^{-1}$, with a fractional systematic error of $1.3\%$.
We discuss in the following a new and more accurate measurement of the same quantity (additional information can be found
in \cite{knote:221}).
\section{Measurement of $e^+e^-\to\pi^+\pi^-$ cross section at DA$\Phi$NE}
\label{subsec:1.2}
The KLOE detector operates at  \DAF, the Frascati \ff, a ``small angle'' \epm\ collider running mainly at a center
of mass energy equal to the $\phi$ meson mass, $W$\ab1020 MeV.
At \DAF, we measure the differential cross section  for $e^+ e^-\to\pi^+\pi^-\gamma$ as a function of the $\pi^+\pi^-$
invariant mass, $M_{\pi\pi}$, for ISR events, and obtain the dipion cross section
$\sigma_{\pi\pi}\equiv\sigma(e^+ e^-\to\pi^+\pi^-)$ from \cite{Binner:1999bt}:
\begin{equation}
s \left. \frac{\dd\sigma(ee\to\pi\pi\gamma)} {\dd M_{\pi\pi}^2}\right|_{{\rm ISR}}=
\sigma_{\pi\pi}(M_{\pi\pi}^2)~ H(M_{\pi\pi}^2,s).
\label{eq:1}
\end{equation}
Eq. \ref{eq:1} defines $H$, the ``radiator function''. $H$ can be obtained from QED calculations and depends on the $e^+e^-$ center-of-mass energy
squared $s$. In Eq. \ref{eq:1} we neglect FSR, which however is
included in our analysis. The cross section we obtain is inclusive of all radiation in the final state.

\section{Selection of $e^+e^-\to\pi^+\pi^-\gamma$ events and background rejection}
\label{sec:2}
KLOE collected a nominal integrated luminosity of about 2.5 fb$^{-1}$ up to the year 2006.
The results presented here use data with $\int\!\L\dif t$=240 pb$^{-1}$ taken in 2002. The
statistical fractional error on the \pic\ contribution to the muon anomaly, \Dam, for dipion masses between 0.592 and 0.975 GeV is smaller than 0.2\%.
\subsection{The KLOE detector}
\label{subsec:1.1}
\begin{figure}
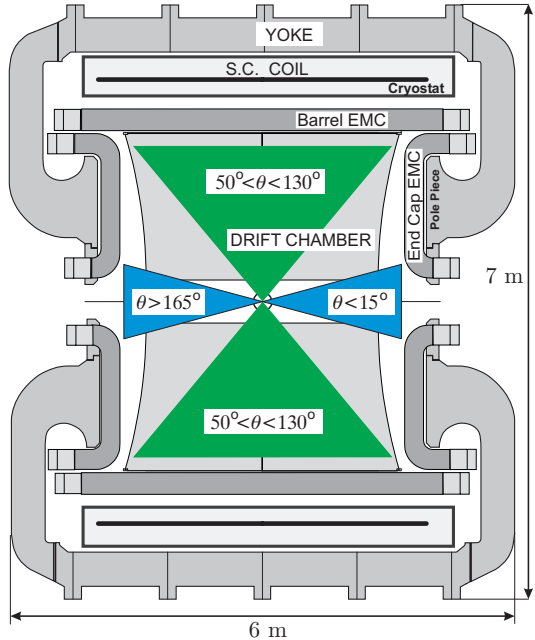

\cl{\lfigbox kloe;7;}
\caption{Vertical cross section of the KLOE detector, showing the small and large angle regions where photons and pions are accepted}
\label{det}
\end{figure}

The KLOE detector consists of a cylindrical drift chamber (DC) \cite{Adinolfi:2002uk}
and an electromagnetic calorimeter (EMC) \cite{Adinolfi:2002zx}.
The DC has a momentum resolution of
$\sigma_{p_\bot}/p_\bot\sim 0.4\%$ for tracks with polar angle $\theta>45^\circ$ .
Track points are measured in the DC with a resolution in $r$-$\phi$ of \ab~0.15 mm and \ab~2 mm in $z$.
The EMC has an energy resolution of $\sigma_E/E\sim 5.7\%/\sqrt{E\ {\rm(GeV)}}$ and an excellent time
resolution of $\sigma_t\sim 54\ {\rm ps}/\sqrt{E\ {\rm(GeV)}}\oplus 100\ {\rm ps}$.
Calorimeter clusters are reconstructed grouping together energy deposits close in space and time.
A superconducting coil provides an axial magnetic field of 0.52 T along the bisector of the colliding beam directions.
The bisector is taken as the $z$ axis of our coordinate system.
The $x$ axis is horizontal, pointing to the center of the collider rings and the $y$ axis is vertical, directed upwards.
A cross section of the detector in the $y,\:z$ plane is shown in Fig. \ref{det}.
Events used in this analysis were triggered \cite{Adinolfi:2002hs}
by two energy deposits larger than 50 MeV
in two sectors of the barrel calorimeter.

The \pic\gam\ cross section diverges at small photon angle as $(1-\cos^2\theta_\gamma)^{-2}\propto1/\theta_\gamma^4$ making
FSR \pic\gam\ processes and $\phi$ decays relatively unimportant. For this reason we measure $\dd\sig/\dd M_{\pi\pi}^2$ at small photon angle.

\begin{figure}[hb!]
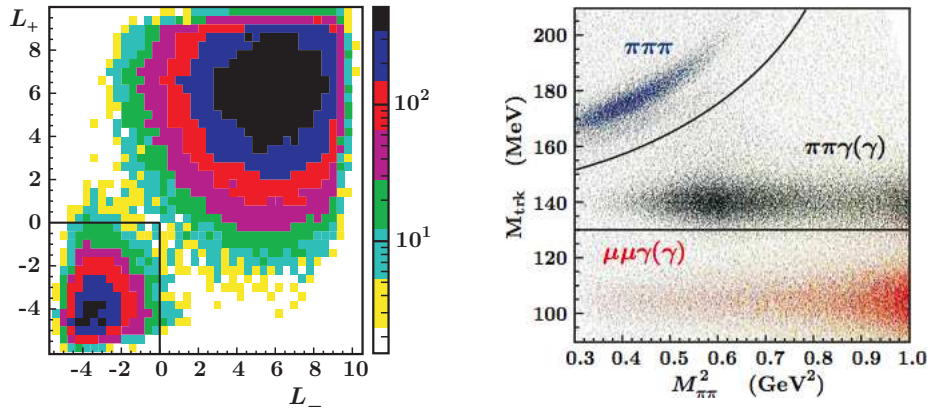

\cl{\lfigbox lpid;5.6;\kern1cm\lfigbox tmass;5.6;}
\caption{Left: PID estimator of the positive track vs. PID estimator of the negative
track. Right: signal and background distributions in the $M_{\rm trk}$--$M^2_{\pi\pi}$ plane.}
\label{mM}
\end{figure}

Figure \ref{det} shows the fiducial volumes we use in the following for pions and photons.
Note that the photon is not detected. Below we list the requirements for event selection.
\begin{enumerate}
\item Two tracks of opposite sign curvature, crossing a cylinder of radius 8 cm and length 15 cm centered at the interaction point,
must satisfy $50^\circ<\theta<130^\circ$ and $p_\bot>160$ MeV or $|p_z|>90$ MeV, to ensure good reconstruction and efficiency.
\item The (unobserved) photon direction, reconstructed from the two tracks above as ${\bf p}_\gam\!=\!-({\bf p}_++ {\bf p}_-)$
must satisfy $|\cos\theta_\gamma|\!>\!\cos(\pi/12)$ (15\deg).
In the following $\theta_{\pi\pi}=\pi-\theta_\gamma$ is the polar angle of the dipion system.
\item The main background processes are $e^+ e^-\to e^+ e^-\gamma,\,\mu^+\mu^-\gamma$ and $\phi\to\pi^+\pi^-\pi^0$ decays.
Signal 
events are distinguished from $e^+ e^-\to e^+ e^-\gamma$ events by particle identification (PID),
using a pseudo-likelihood estimator \cite{knote:192} for each track, $L_\pm$,
based on time of flight, energy and shape of the cluster associated to the track. 
Electrons deposit most of their energy near the entrance of the calorimeter while muons and pions lose energy almost uniformly
along the depth of the calorimeter.
Events with both tracks having $L_\pm<0$, as in the lower left rectangle of Fig. \ref{mM},
left, are identified as electrons and rejected. The efficiency for this selection is
larger than $99.95\%$, evaluated on $\pi^+\pi^-\gamma$ samples, obtained from both data and Monte Carlo.
The probability for the $e^+e^-\gamma$ events to be misidentified by the estimator as $\pi^+\pi^-\gamma$ is $3\%$.
\item The event must satisfy a cut on the track mass variable, $M_{\rm trk}$. Assuming the presence of an unobserved photon
and that the tracks belong to particles of the same mass, $M_{\rm trk}$ is computed from energy and momentum conservation:
$$
\left(\sqrt{s}-\sqrt{|\mathbf{p_+}|^2 + M^2_{\rm trk}}-
\sqrt{|\mathbf{p_-}|^2 + M^2_{\rm trk} }\right)^2-\left(\mathbf{p_+}
+\mathbf{p_-}\right)^2 = 0
$$
where $\mathbf{p_\pm}$ is the measured momentum of the positive (negative) particle,
and only one of the four solutions is physical.
A cut is applied in the $M_{\rm trk}$--$M^2_{\pi\pi}$ plane,
as shown in Fig. \ref{mM}. The requirement $M_{\rm trk}>130\MeV$ rejects $\mu^+\mu^-\gamma$,
and further suppresses the fraction of $e^+ e^-\gamma$ surviving the pseudo-likelihood
selection. 
While the $M^2_{\pi\pi}$ dependent curve rejects $\pi^+\pi^-\pi^0$.
\end{enumerate}
About $3\times10^6$ events pass these criteria.


Residual \eeg\, , \mmg\ and \ppp\ background contamination levels are evaluated by fitting the $M_{\rm trk}$ spectrum of the accepted
events with a superposition of Monte Carlo (MC) distributions describing signal and backgrounds.
The normalized contributions from 
signal and backgrounds are free parameters of the fit, performed for 30 intervals in $M_{\pi\pi}^2$ of 0.02 GeV\up2 width.
We also estimate the contribution from \epm\to\pic\epm events, using the MC generator Ekhara \cite{Czyz:2006dm}.
The errors on the background estimates reflect the uncertainty on the production mechanism and the errors on the
normalization coefficients from the fit. The results are given in Table \ref{tab:syseff_bkg}.
Other possible backgrounds from \cite{Pancheri:2007xt}
$\phi\to(f_0,\sigma)\gamma\to\pi^+\pi^-\gamma$, $e^+e^-\to\rho^{\pm}\pi^{\mp}\to\pi^+\pi^-\gamma$,
$\phi\to(\eta,\eta')\gamma$ and the radiative return to the $\omega$, $e^+e^-\to\omega\gamma_{\rm ISR}\to\pi^+\pi^-\pi^0\gamma$, are negligible
after the acceptance cuts. Background from the process $e^+e^-\to e^+e^-e^+e^-$ is estimated to be well below 0.1\%.

\subsection{Improvements with respect to the published analysis}
\label{subsec:1.3}
With respect to our published result \cite{Aloisio:2004bu},
the present analysis profits from lower machine background and more stable \DAF\ operation in 2002.
Improved data filters were also developed.
\begin{enumerate}
\item A new trigger (L3) implemented at the end of 2001 reduced the loss of events rejected as cosmic rays events
from 30\% to 0.2\%.
The event loss is determined from a downscaled control sample taken without the enforcement of the cosmic-ray veto.
\item An offline background filter efficiency of 95\% resulted in a large systematic uncertainty in our previous result.
A new filter with 98.5\% efficiency and negligible systematic uncertainty has been implemented.
A downscaled sample is retained to evaluate the filter efficiency.
\item We no longer require that the two pion tracks form a vertex. The uncertainty
in the corresponding efficiency in the 2005 analysis is now removed.
\end{enumerate}
Finally, the Bhabha cross section has been reevaluated with better accuracy\cite{Balossini:2006wc}
than at the time of our previous measurement, as discussed below.

\subsection{Efficiencies, acceptance and systematic errors}
\label{subsec:1.4}
An improved simulation of the detector response allows
determination of the efficiency for the event-selection
criteria described above. The 
generator Phokha\-ra, including next-to-leading-order ISR \cite{Rodrigo:2001kf} and leading-order FSR corrections,
as well as simultaneous emission of one ISR and one FSR photon \cite{Czyz:2003ue}
has been inserted in the standard KLOE MC Geanfi \cite{Ambrosino:2004qx}.

Corrections are needed for the trigger and tracking efficiency. 
We compare MC efficiencies with efficiencies obtained from a data control sample, and
correct the MC where discrepancies are present.

{\bf Trigger.}\kern3mm The efficiency is obtained from
a subsample of $\pi^+\pi^-\gamma$ events in which a single pion
satisfies the trigger requirements. Then, the trigger response
for the other pion is parametrized as a function of its momentum and
direction.
The efficiency
as a function of $M_{\pi\pi}$ is obtained using the MC event distribution
and ranges from 96\% to 99\%.
The result is checked with a subset of the same sample, selected with an essentially independent drift
chamber trigger, and evaluating the efficiency satisfying the calorimeter trigger
directly as a function of $M_{\pi\pi}$. The constant fractional difference of 0.1\% is taken as the systematic uncertainty.

{\bf Tracking.}\kern3mm The tracking efficiency for single pions is evaluated for each charge as a function of momentum
and polar angle, using \f\to\pic\po\ and \pic\gam\ events, both identified
on the basis of calorimeter information and the observation of a pion track.
The two control samples are complementary: \f\to\pic\po\ decays are more abundant, but do not cover the whole momentum range of interest.
The efficiency is \ab~98\% and constant in $M_{\pi\pi}^2$.
The fractional difference between the results obtained with the two samples is 0.3\%, which is taken as systematic error.

{\bf Pion ID.}\kern3mm
Each track is extrapolated  to the calorimeter and at least one cluster is searched for within a sphere of radius
$|{\bf r}_{\rm ent}\!-\!{\bf r}_{\rm clu}|\!<\!$ 90 cm.
$\mathbf{r}_{\rm ent}$ and $\mathbf{r}_{\rm clu}$ are the coordinates of the track entry point
and of the cluster centroid.  The value of 90 cm is chosen to include pion fragments and
minimizes the systematic error on the trigger efficiency correction.
The efficiency is evaluated on $\pi^+\pi^-\gamma$ events where a single track
has an associated cluster identified as pion ($L>0$), and parametrizing the probability
of the other track to find an associated cluster with $L>0$. This probability is larger
than 99.9\%.
No difference on the pion cluster efficiency is found by
varying the association radius; the systematic error is taken as negligible.

\begin{figure}[ht]
\centering
\epsfig{file=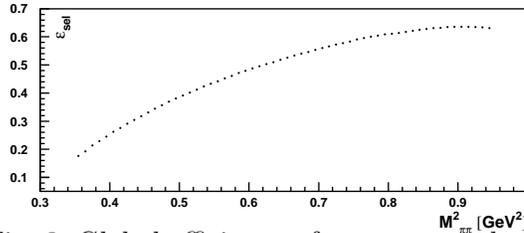,width=8.1cm}
\vglue-0.55cm
\caption{Global efficiency of $\pi\pi\gamma$ event selection.}
\label{eff_glob}
\end{figure}

Efficiencies for $M_{\rm trk}$ cuts and acceptance are
evaluated from MC, corrected to reproduce data distributions. 
The efficiency of the selection is shown in Fig. \ref{eff_glob}.

Systematic uncertainties are obtained as follows.
\begin{itemize}
\item The systematic uncertainty due to the $M_{\rm trk}$ cut is obtained
by moving the cuts in the $M_{\rm trk}$--$M^2_{\pi\pi}$ plane within reasonable limit of the mass resolution and
evaluating the difference in the $\pi^+\pi^-\gamma$ spectrum.
We find a fractional difference of 0.2\% (constant in $M_{\pi\pi}$)
which we take as systematic error.
\item Systematic effects due to polar angle requirements for the pions, $50\deg<\theta<130\deg$, and of dipion, $|\cos\theta_{\pi\pi}|\!>\!\cos(15\deg)$,
are estimated by varying the angular acceptance by $1^\circ$
around the nominal value. The systematic error
decreases from 0.6\% at $M_{\pi\pi}^2=0.35 \GeV^2$ to 0.1\% at $M_{\pi\pi}^2=0.64 \GeV^2$
 and is negligible for larger values (see Table \ref{tabacc}).
\end{itemize}

\subsection{Luminosity}
\label{subsec:1.5}
The absolute normalization of the data sample is obtained from the yield of large angle ($55\deg\!<\!\theta\!<\!125\deg$) Bhabha-scattering events.
The integrated luminosity, $\int\!\L\dif t$, is obtained by dividing the observed number of these events by the effective cross section
evaluated by the MC generator
Babayaga \cite{Carloni Calame:2000pz}, 
inserted in the KLOE MC Geanfi \cite{Ambrosino:2006te}.
The Babayaga generator includes QED radiative corrections via the parton shower algorithm.
An updated version of the generator, Babayaga@NLO \cite{Balossini:2006wc}, gives a Bhabha cross section of 456.2 nb,
0.7\% lower than the value from the previous version, while the theoretical
uncertainty is reduced from 0.5\% to 0.1\%. The experimental uncertainty on the luminosity is 0.3\%, dominated by
the systematics on the angular acceptance.

\section{Evaluation of the pion form factor}
\label{sec:3}

The differential $\pi^+\pi^-\gamma$ cross section is obtained from the observed number of events,
$N_{\rm obs}$, after subtracting the residual background, $N_{\rm bkg}$, correcting for
the selection efficiency, $\epsilon_{\rm sel}(M_{\pi\pi}^2)$, and dividing by the luminosity \L:
\begin{equation}
\frac{\dd\sigma_{\pi\pi\gamma}}
{\dd M_{\pi\pi}^2} = \frac{N_{\rm obs}-N_{\rm bkg}}
{\Delta M_{\pi\pi}^2}\, \frac{1}{\epsilon_{\rm sel}(M_{\pi\pi}^2)~ \mathcal{L}}.
\label{eq:2}
\end{equation}
The background mentioned above varies smoothly from 1\% around the $\rho$ peak to \ab7\%, for low and
high $M_{\pi\pi}$ mass. This background is dominated by misidentified muon pairs at high mass, and $\pi^+\pi^-\pi^0$
events at low mass.
Our mass resolution ($\delta M_{\pi\pi}^2\sim 2\times 10^{-3} \GeV^2$, or $\delta M_{\pi\pi}$\ab1.3 MeV at the $\rho$ peak) allows 
us to use bins of $\Delta M_{\pi\pi}^2=0.01\GeV^2$ width.
In order to correct for resolution effects,
the differential cross section is unfolded using the Bayesian method described in \cite{D'Agostini:1994zf}.
The point-by-point uncertainty introduced by the unfolding procedure -- appreciable only in the $\rho$-$\omega$
region, $M_{\pi\pi}^2\sim0.6\GeV^2$ --
is given in Table \ref{tab:syseff4}.
The unfolding does not introduce any additional systematic error on \Dam.

The quantity $M_{\pi\pi}$ is computed from measured momenta of the pions and is shifted by radiative
effects from the mass value at the $\pi^+\pi^-\gamma$ vertex, $M^0_{\pi\pi}$.
The cross section $\sigma_{\pi\pi}(M^0_{\pi\pi})$ is determined accounting for this shift
and dividing the $\pi^+\pi^-\gamma$ cross section by the radiator function $H$
(obtained from Phokhara by setting pion form factor $F_\pi=1$) as in Eq. \ref{eq:1}.
Fluctuations in the $e^+e^-$ CM energy during data taking
introduce an additional systematic uncertainty in the evaluation of $H$
(see Table \ref{tab:syseff_herr}).

The pion form factor is obtained from $\sigma_{\pi\pi}$ after subtraction of FSR,
assuming point-like pions ($\eta_{FSR}$ term \cite{Schwinger:1989ix,Jegerlehner:2006ju}):
\begin{equation}
|F_\pi|^2\left(1+\eta_{\rm FSR}\right) = \frac{3}{\pi}\:\frac{M^2_{\pi\pi}}{\alpha^2\:\beta_\pi^3}\,\sigma_{\pi\pi}~,~ ~
\beta_\pi = \sqrt{1 - \frac{4\,m_\pi^2}{M^2_{\pi\pi}}} 
\end{equation}
where $\alpha$ is the fine structure constant ($\alpha=e^2/4\pi$), and $\beta_\pi$ is the pion velocity
in the $\pi\pi$ rest frame.

Our results are summarized in Table \ref{tab:1},
which gives the integrals over bins of 0.01 GeV\up2 identified by their lower edge of:
\begin{itemize}
\item the observed differential cross section $\dd\sigma_{e^+ e^- \to \pi\pi\gamma}/\dd M_{\pi\pi}^2$
as a function of the invariant mass of the dipion system in the angular region $\theta_{\pi\pi}\, (\pi-\theta_{\pi\pi})< 15^\circ$;
\item the {\it bare} cross section $\sigma^0_{\pi\pi}$, inclusive of FSR,
and with vacuum polarization effects removed \cite{vacpol}:
$\sigma_{\pi\pi}^0=\sigma_{\pi\pi}\left[\alpha(0)/\alpha(M_{\pi\pi})\right]^2$;
\item the pion form factor without FSR and with vacuum polarization effects included.
\end{itemize}
The errors in Table \ref{tab:1}
are statistical only. The systematic errors are given in Tables \ref{tab:syseff_bkg},
\ref{tabacc}, \ref{tab:syseff4}, and \ref{tab:syseff_herr}.
The statistical errors are weakly correlated as a result of the resolution unfolding.
The correct covariance matrix is used for the calculation of the statistical error on \Dam.
The systematic errors cannot be considered as fluctuations of random variables within each of the categories considered.
We combine all contributions for the same $M_{\pi\pi}$ value in quadrature.
We add linearly then the errors for each bin to obtain the total systematic error on \Dam.

\begin{figure}[ht!]
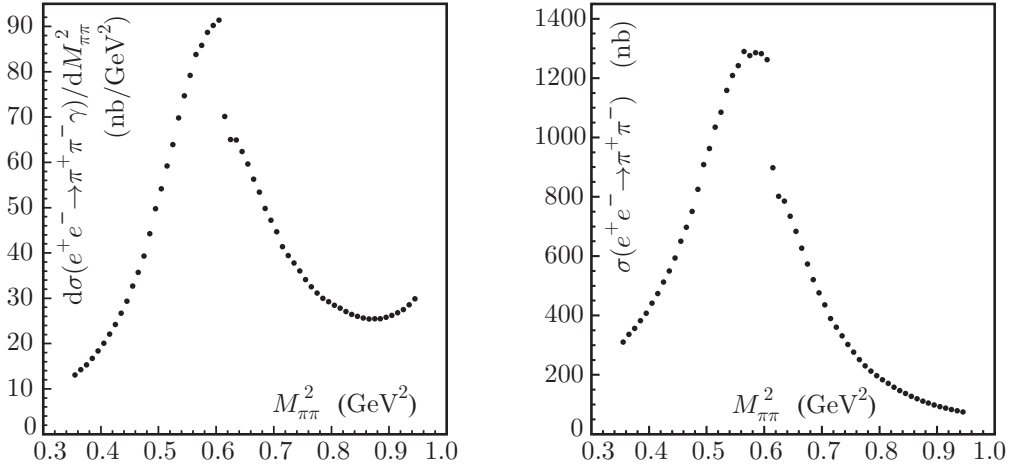
\centering
\lfigbox sppg;6;\kern1cm\lfigbox spp;6.3;
\caption{Left: Differential cross section for \epm\to\pic\gam(\gam),
with $|\cos\theta_\gamma|>\cos(15\deg)$. Right: cross section for $e^+e^-\to \pi^+\pi^-$.}
\label{fig:sigppg}
\end{figure}

Figure \ref{fig:sigppg} left, shows the differential cross section for \epm\to\pic\gam~with
$|\cos\theta_\gamma|>\cos(15\deg)$ after applying the corrections described above.
Figure \ref{fig:sigppg} right, shows the cross section $\sigma^0_{\pi\pi}$,
which is the input for the dispersive integral for \Dam.
\begin{table}[htbp]
\begin{center}
\renewcommand{\arraystretch}{.9}
\begin{tabular}{||c|c|c|c|c|c|c|c|c|c|c||}
\hline\hline\noalign{\vglue.5mm}
$M^2_{\pi\pi}$ (GeV$^2$) & 0.00 & 0.01 & 0.02 & 0.03 & 0.04 & 0.05 & 0.06 & 0.07 & 0.08 & 0.09 \\
\hline
0.3&  &  &  &  &  &0.5&0.4&0.4&0.4&0.4\\
0.4&0.4&0.4&0.4&0.3&0.3&0.3&0.3&0.3&0.3&0.3\\
0.5&0.3&0.3&0.3&0.2&0.3&0.2&0.2&0.2&0.2&0.2\\
0.6&0.2&0.2&0.2&0.2&0.2&0.2&0.2&0.2&0.2&0.2\\
0.7&0.2&0.1&0.1&0.1&0.1&0.1&0.1&0.1&0.1&0.1\\
0.8&0.1&0.1&0.1&0.1&0.1&0.1&0.1&0.1&0.1&0.1\\
0.9&0.1&0.1&0.1&0.1&0.1& & & & & \\

\hline\hline
\end{tabular}\vglue2mm
\caption{Systematic errors in (in percent) 
due to background subtraction, in
0.01 GeV$^2$ bin intervals of $M^2_{\pi\pi}$. The lower bin's edge
is given by the sum of the values in the first row and first column.}
\label{tab:syseff_bkg}
\end{center}
\end{table}
\begin{table}[htp]
 \centering
 \renewcommand{\arraystretch}{1.}
 \setlength{\tabcolsep}{0.9mm}
\begin{tabular}{||l|c||}
\hline\hline
$M_{\pi\pi}^2$ range ($\GeV^2$)  & Systematic error (\%) \\
\hline
0.35 $\leq M_{\pi\pi}^2 < 0.39$  & 0.6 \\
0.39 $\leq M_{\pi\pi}^2 < 0.43$  & 0.5 \\
0.43 $\leq M_{\pi\pi}^2 < 0.45$  & 0.4 \\
0.45 $\leq M_{\pi\pi}^2 < 0.49$  & 0.3 \\
0.49 $\leq M_{\pi\pi}^2 < 0.51$  & 0.2 \\
0.51 $\leq M_{\pi\pi}^2 < 0.64$  & 0.1 \\
0.64 $\leq M_{\pi\pi}^2 < 0.95$  & - \\
\hline\hline
\end{tabular}
\vglue-0.1cm
\vspace{0.4cm}
\caption{Fractional systematic errors on the acceptance.}
\label{tabacc}
\end{table}
\begin{table}[htb]
\centering
\renewcommand{\arraystretch}{1.15}
\begin{tabular}{||c|c|c|c|c|c||}
\hline\hline
$M_{\pi\pi}^2$ (GeV$^2$) &  0.58 & 0.59 & 0.6 & 0.61 & 0.62 \\
\hline
$\delta_{unf} (\%)$ & 0.4 & 0.3 & 2.1 & 4.0 & 0.4  \\
\hline\hline
\end{tabular}
\caption{Systematic error (in percent) on
\mbox{$d\sigma_{e^+e^-\to\pi\pi\gamma}/dM_{\pi\pi}^2$} due to
the correction for detector resolution in 0.01 GeV$^2$ intervals. The
indicated values for $M_{\pi\pi}^2$ represent the lower bin's edge. Outside this interval,
the $\rho$-$\omega$ region, the effect is negligible.}
\label{tab:syseff4}
\end{table}
\begin{table}[ht]
\begin{center}
\renewcommand{\arraystretch}{.9}
\begin{tabular}{||c|c|c|c|c|c|c|c|c|c|c||}
\hline\hline\noalign{\vglue.5mm}
$(M^0_{\pi\pi})^2$ (GeV$^2$) & 0.00 & 0.01 & 0.02 & 0.03 & 0.04 & 0.05 & 0.06 & 0.07 & 0.08 & 0.09 \\\hline
0.3&  &  &  &  &  &0.3&0.4&0.2&0.1&0.2\\
0.4&0.1&0.2&0.1&0.2&0.1&0.2&0.2&0.3&0.1&0.2\\
0.5&0.3&0.3&0.1&0.2&0.3&0.3&0.3&0.2&0.2&0.3\\
0.6&0.1&0.3&0.3&0.2&0.2&0.3&0.3&0.3&0.2&0.3\\
0.7&0.3&0.3&0.4&0.3&0.3&0.3&0.4&0.3&0.3&0.3\\
0.8&0.4&0.4&0.3&0.5&0.4&0.5&0.4&0.4&0.5&0.5\\
0.9&0.5&0.6&0.6&0.6&0.7& & & & & \\

\hline\hline
\end{tabular}\vglue2mm
\caption{Systematic errors (in percent) on the radiator function
due to the spread of $\sqrt{s}$ in the 2002 data taking period, given in
0.01 GeV$^2$ bin intervals of $M^2_{\pi\pi}$. The lower bin's edge
is given by the sum of the values in the first row and first column.}
\label{tab:syseff_herr}
\end{center}
\end{table}
\begin{table}
\begin{center}
\renewcommand{\arraystretch}{1.0}
\begin{tabular}{||l|c|c|c|c||}
\hline
 & $\sigma_{\pi\pi\gamma}$ & $\sigma_{\pi\pi}^0$ & $F_\pi$ & \Dam \\
\hline
\hline
Reconstruction Filter & \multicolumn{4}{|c||}{negligible} \\
\cline{2-5}
Background subtraction & \multicolumn{3}{|c|} {Tab. \ref{tab:syseff_bkg}} & 0.3\% \\
\cline{2-5}
Trackmass & \multicolumn{4}{|c||}{0.2\%} \\
\cline{2-5}
Pion cluster ID & \multicolumn{4}{|c||}{negligible} \\
\cline{2-5}
Tracking efficiency & \multicolumn{4}{|c||}{0.3\%} \\
\cline{2-5}
Trigger efficiency & \multicolumn{4}{|c||}{0.1\%}\\
\cline{2-5}
Acceptance & \multicolumn{3}{|c|}{Tab. \ref{tabacc}} & 0.2\%\\
\cline{2-5}
Unfolding & \multicolumn{3}{|c|}{Tab. \ref{tab:syseff4}} & {negligible} \\
\cline{2-5}
L3 filter & \multicolumn{4}{|c||}{0.1\%} \\
\cline{2-5}
$\sqrt{s}$ dependence of $H$ & - & \multicolumn{2}{|c|}{Tab. \ref{tab:syseff_herr}} & 0.2\% \\
\cline{2-5}
Luminosity & \multicolumn{4}{|c||}{0.3\%} \\
\cline{2-5}\hline
Experimental systematics & \multicolumn{3}{|c|}{ } & 0.6\% \\\hline\hline
FSR resummation & - & \multicolumn{3}{|c||}{0.3\%} \\
\cline{2-5}
Radiator function $H$ & - & \multicolumn{3}{|c||}{0.5\%} \\
\cline{2-5}
Vacuum Polarization & - & 0.1\% & - & 0.1\% \\
\cline{2-5}\hline
Theory systematics & \multicolumn{3}{|c|}{ } & 0.6\% \\
\hline\hline
\end{tabular}
\caption{Systematic errors on $\sigma_{\pi\pi\gamma}$, $\sigma_{\pi\pi}^0$, $F_\pi$ and \Dam.}
\label{tab:syseff3}
\end{center}
\end{table}

\section{Evaluation of $\Delta^{\pi\pi}a_\mu$}
The dispersive integral for \Dam\ is computed as the sum of the values for $\sigma_{\pi\pi}^0$ listed
in Table \ref{tab:1} times the kernel $K(s)$:
\begin{equation}
\Dam=\frac{1}{4\pi^3}\int_{s_{min}}^{s_{max}}\dd s\,\sigma_{\pi\pi(\gamma)}^0(s)\,K(s) ~,
\label{amuint}
\end{equation}
where the kernel, see the second paper of ref. \cite{deraf}, is given by
$$K(s)= x^2\Bigl(1-{x^2\over2}\Bigr) 
+(1+x)^2(1+x^{-2}) \Bigl(\log(1+x)-x+{x^2\over2}\Bigr) 
+{1+x\over1-x}x^2\log x 
$$
with
$$x={1-\sqrt{1-4m_\mu^2/s}\over1+\sqrt{1-4m_\mu^2/s}}$$

Eq. \ref{amuint} gives
$\Dam =  (387.2\pm0.5_{\rm stat}\pm2.4_{\rm exp}\pm2.3_{\rm th})\times10^{-10}$ in the interval 0.35$<M_{\pi\pi}^2<$ and 0.95 GeV\up2.
Contributions to the systematic errors on \Dam\ are given in the last column of Table \ref{tab:syseff3}.

\section{Comparison between 2008 and 2005 analyses}
In order to compare consistently the $\pi^+\pi^-\gamma$ differential cross section from
this analysis to that from our previous analysis,
two corrections have been applied to the previous results:
\begin{itemize}
\item a $-0.7\%$ overall shift, due to the new evaluation of the Bhabha cross section,
obtained from the updated version of the Babayaga generator (see Sec. \ref{subsec:1.5});
\item an energy-dependent effect due to a
double counting of the calorimeter cluster efficiency
in the evaluation of the trigger correction, which overestimates the cross section
mainly at low-mass values by a few percent.
\end{itemize}
\begin{figure}[ht]
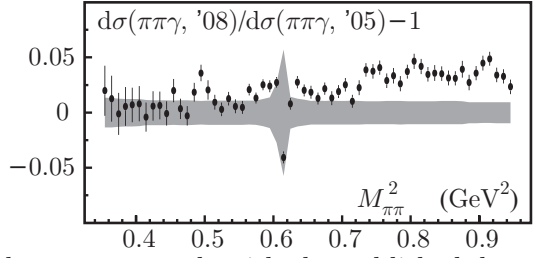

\centering
\lfigbox 08vs05;7;
\vglue-0.5cm
\caption{Comparison of the present result with the published data, updated for the effects described in the text.
The band is just the fractional systematic error of the ratio.} 
\label{01vs02}
\end{figure}

As a result of these updates, the value of \Dam\ from our previous analysis changes to 
$(384.4 \pm~0.8_{stat} \pm~4.6_{sys})\times10^{-10}$.
The fractional difference between the spectra for the present analysis and that previously
published (with updates), is shown in Fig. \ref{01vs02}.
While the agreement below the $\rho$ peak is good, above 0.7 $\GeV^2$ there is some
difference between the two spectra.
The value obtained for the integral
is consistent between the two data sets (as shown in Table \ref{tab:2}).
Because of the improvements to the analysis described in Sec. \ref{subsec:1.3}
and the high quality of the 2002 data, we consider the present result
to supersede that previously published.
\begin{table}[hb]
\centering
 \renewcommand{\arraystretch}{1.1}
 \setlength{\tabcolsep}{1.2mm}
\begin{tabular}{|l|c|}
\hline
\multicolumn{2}{|c|}{\Dam\x10\up{10} \ $0.35<M_{\pi\pi}^2<0.95\GeV^2$}\\\hline
published 05 & $388.7 \pm~0.8_{\rm stat} \pm~4.9_{\rm sys}$\\\hline
updated 05 & $384.4 \pm~0.8_{\rm stat} \pm~4.6_{\rm sys}$\\\hline
new data 08 & $387.2 \pm~0.5_{\rm stat} \pm~3.3_{\rm sys}$\\\hline
\end{tabular}
\caption{\label{tab:2}Comparison among \Dam\ values from KLOE analyses.}
\end{table}

\section{Comparison with CMD-2 and SND results}
\begin{figure}[ht!]
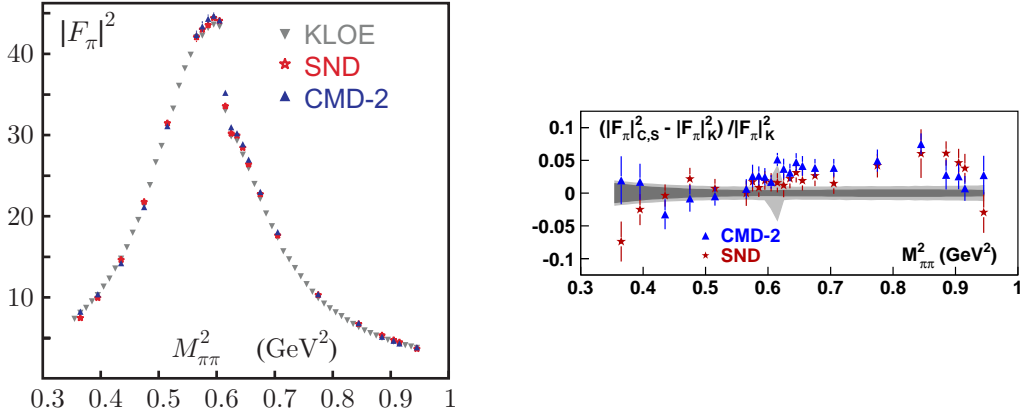

\cl{\lfigbox  fpi;6;\kern1cm\lfigbox dfpi;6.5;}
\caption{Left. $|F_\pi|^2$ from CMD-2 \cite{Akhmetshin:2006bx},
  SND \cite{Achasov:2006vp} and KLOE with statistical errors. Right. Fractional difference between CMD-2 (C) or SND (S) and the
KLOE (K) results. The dark (light) band is the KLOE statistical (statistical $\oplus$ systematic) error.
Also for CMD-2 and SND statistical $\oplus$ systematic errors are shown.}
\label{fig:fpi_kloe_novo}
\end{figure}

We may compare the present result on $|F_\pi|^2$ with the results
from the energy scan experiments at Novosibirsk
CMD-2 \cite{Akhmetshin:2006bx} and SND\cite{Achasov:2006vp}.
For a given energy scan experiment, whenever there are several data points falling in one
$0.01\GeV^2$ bin, we average the values.
The result can be seen in Fig. \ref{fig:fpi_kloe_novo}, left.
Figure \ref{fig:fpi_kloe_novo}, right, shows the fractional difference between
the data points from the energy scan experiments (CMD-2 and SND)
and the KLOE data.
There is reasonable agreement between the experiments, as also indicated by the computed values of $\Delta^{\pi\pi}a_\mu$
given below in the range of overlap 0.630$<M_{\pi\pi}<0.958\GeV$, combining statistical and systematic errors in quadrature:
$$\eqalign{
\hbox{SND,  2006 \cite{Achasov:2006vp}}\ \ \Delta^{\pi\pi}a_\mu&=(361.0\pm5.1)\x10^{-10}\cr
\hbox{CMD-2,  2007 \cite{Akhmetshin:2006bx}}\ \ \Delta^{\pi\pi}a_\mu&= (361.5\pm3.4)\x10^{-10}\cr
\hbox{this work}\ \ \Delta^{\pi\pi}a_\mu&=(356.7\pm3.1)\x10^{-10}.\cr}$$
A fit for the best value gives 359.2\plm2.1 with $\chi^2$/dof=1.24/2, corresponding to a confidence level of 54\%.
\section{Conclusions}
\label{sec:6}
We have measured the dipion contribution to the muon anomaly, $\Delta^{\pi\pi}a_\mu$, in the interval $0.592\km\,<\km\, M_{\pi\pi}\km\,<\km\,0.975$ GeV, with negligible statistical error and a 0.6\% experimental systematic uncertainty.
Radiative corrections calculations increase the systematic uncertainty to 0.9\%. Combining all errors we find:
$$\Dam(0.592\km\,<\km\, M_{\pi\pi}\km\,<\km\,0.975\ {\rm GeV})=(387.2\plm3.3)\x10^{-10}.$$
This result is consistent with our previous value, with a total error smaller by 30\%.
Our new result confirms the current disagreement between the standard model prediction for $a_\mu$ and
the measured value.

\section*{Acknowledgements}
We would like to thank Carlo Michel Carloni Calame, Henryk Czy{\.z},
Fred Jegerlehner, Johann K\"uhn, Guido Montagna, Germ{\'a}n Rodrigo, and Olga Shekhovtsova
for numerous useful discussions.
We thank the DA$\Phi$NE team for their efforts in maintaining low background running
conditions and their collaboration during all data-taking.
We want to thank our technical staff:
G.F. Fortugno and F. Sborzacchi for ensuring the efficient operation of the KLOE computing facilities; M.~Anelli for his continuous attention to the gas system and the safety of the detector;
A.~Balla, M.~Gatta, G.~Corradi and G.~Papalino for maintenance of the electronics;
M.~Santoni, G.~Paoluzzi and R.~Rosellini for the general support to the detector;
C.~Piscitelli for his help during major maintenance periods.
This work was supported in part by EURODAPHNE, contract FMRX-CT98-0169;
by the German Federal Ministry of Education and Research (BMBF) contract 06-KA-957;
by the German Research Foundation (DFG), 'Emmy Noether Programme', contracts DE839/1-4;
by INTAS, contracts 96-624, 99-37; and by the EU Integrated Infrastructure Initiative HadronPhysics Project under contract number RII3-CT-2004-506078.
\vglue3mm
\begin{table}[ht]
\centering
\renewcommand{\arraystretch}{1.}
\renewcommand{\tabcolsep}{3mm}
{\small
\begin{tabular}{|c|c|c|c||c|c|c|c|}
\hline
\kma$M^2_{\pi\pi}$\km&\kern-2.5mm$\dd\sigma_{\pi\pi\gam}/\dd M^2_{\pi\pi}$\kern-3mm&$\sigma^0_{\pi\pi(\gamma)}$&\parbox{1cm} {\vglue3mm$|F_\pi|^2$\vglue-3mm}&%
\kma$M^2_{\pi\pi}$\km&\kern-2.5mm$\dd\sigma_{\pi\pi\gam}/\dd M^2_{\pi\pi}$\kern-3mm&$\sigma^0_{\pi\pi(\gamma)}$&\parbox{1cm} {\vglue3mm$|F_\pi|^2$\vglue-3mm}\\
\kma GeV$^2$\kma& nb/GeV$^2$    &   nb            &    &
\kma GeV$^2$\kma& nb/GeV$^2$    &   nb            &    \\
 \hline
\km0.355\km\kak13.07\plm0.16\km\kak309\plm4\km\kak7.35\plm0.11\kma&%
\km0.655\km\kak59.62\plm0.19\km\kak683.8\plm2.7\km\kak25.90\plm0.10\kma\\
\km0.365\km\kak14.21\plm0.16\km\kak335\plm4\km\kak8.09\plm0.11\kma&%
\km0.665\km\kak56.28\plm0.18\km\kak626.9\plm2.5\km\kak23.98\plm0.10\kma\\
\km0.375\km\kak15.20\plm0.16\km\kak354\plm4\km\kak8.68\plm0.11\kma&%
\km0.675\km\kak53.43\plm0.18\km\kak573.5\plm2.4\km\kak22.16\plm0.09\kma\\
\km0.385\km\kak16.60\plm0.16\km\kak380\plm4\km\kak9.45\plm0.11\kma&%
\km0.685\km\kak49.84\plm0.17\km\kak520.8\plm2.2\km\kak20.33\plm0.09\kma\\
\km0.395\km\kak18.23\plm0.17\km\kak405\plm4\km\kak10.23\plm0.11\kma&%
\km0.695\km\kak47.22\plm0.16\km\kak476.0\plm2.0\km\kak18.78\plm0.08\kma\\
\km0.405\km\kak19.97\plm0.16\km\kak439\plm4\km\kak11.28\plm0.11\kma&%
\km0.705\km\kak44.65\plm0.16\km\kak435.8\plm1.9\km\kak17.38\plm0.08\kma\\
\km0.415\km\kak22.00\plm0.17\km\kak472\plm4\km\kak12.30\plm0.11\kma&%
\km0.715\km\kak41.40\plm0.15\km\kak389.5\plm1.7\km\kak15.70\plm0.07\kma\\
\km0.425\km\kak24.09\plm0.17\km\kak511\plm4\km\kak13.51\plm0.11\kma&%
\km0.725\km\kak39.40\plm0.14\km\kak360.7\plm1.6\km\kak14.69\plm0.07\kma\\
\km0.435\km\kak26.57\plm0.17\km\kak548\plm4\km\kak14.70\plm0.11\kma&%
\km0.735\km\kak37.80\plm0.14\km\kak331.1\plm1.5\km\kak13.63\plm0.06\kma\\
\km0.445\km\kak29.26\plm0.18\km\kak592\plm4\km\kak16.13\plm0.12\kma&%
\km0.745\km\kak36.05\plm0.14\km\kak302.6\plm1.4\km\kak12.60\plm0.06\kma\\
\km0.455\km\kak32.56\plm0.19\km\kak648\plm4\km\kak17.91\plm0.12\kma&%
\km0.755\km\kak34.13\plm0.13\km\kak276.0\plm1.3\km\kak11.63\plm0.05\kma\\
\km0.465\km\kak35.60\plm0.19\km\kak695\plm4\km\kak19.49\plm0.12\kma&%
\km0.765\km\kak32.50\plm0.13\km\kak251.4\plm1.2\km\kak10.70\plm0.05\kma\\
\km0.475\km\kak39.18\plm0.19\km\kak749\plm4\km\kak21.31\plm0.13\kma&%
\km0.775\km\kak31.14\plm0.12\km\kak230.2\plm1.1\km\kak9.91\plm0.05\kma\\
\km0.485\km\kak44.28\plm0.20\km\kak826\plm5\km\kak23.85\plm0.13\kma&%
\km0.785\km\kak30.01\plm0.12\km\kak212.3\plm1.0\km\kak9.24\plm0.04\kma\\
\km0.495\km\kak49.73\plm0.21\km\kak908\plm5\km\kak26.61\plm0.14\kma&%
\km0.795\km\kak29.23\plm0.11\km\kak197.4\plm0.9\km\kak8.68\plm0.04\kma\\
\km0.505\km\kak54.17\plm0.22\km\kak963\plm5\km\kak28.65\plm0.14\kma&%
\km0.805\km\kak28.46\plm0.11\km\kak183.7\plm0.9\km\kak8.16\plm0.04\kma\\
\km0.515\km\kak59.20\plm0.22\km\kak1035\plm5\km\kak31.25\plm0.15\kma&%
\km0.815\km\kak27.79\plm0.11\km\kak171.3\plm0.8\km\kak7.69\plm0.04\kma\\
\km0.525\km\kak63.90\plm0.23\km\kak1085\plm5\km\kak33.25\plm0.15\kma&%
\km0.825\km\kak27.06\plm0.11\km\kak158.4\plm0.8\km\kak7.180\plm0.035\kma\\
\km0.535\km\kak69.82\plm0.24\km\kak1158\plm5\km\kak36.05\plm0.16\kma&%
\km0.835\km\kak26.43\plm0.10\km\kak147.0\plm0.7\km\kak6.732\plm0.032\kma\\
\km0.545\km\kak74.68\plm0.24\km\kak1209\plm5\km\kak38.22\plm0.16\kma&%
\km0.845\km\kak26.02\plm0.10\km\kak137.5\plm0.6\km\kak6.358\plm0.030\kma\\
\km0.555\km\kak79.20\plm0.24\km\kak1242\plm5\km\kak39.88\plm0.16\kma&%
\km0.855\km\kak25.63\plm0.10\km\kak127.4\plm0.6\km\kak5.948\plm0.028\kma\\
\km0.565\km\kak83.79\plm0.25\km\kak1289\plm5\km\kak42.06\plm0.16\kma&%
\km0.865\km\kak25.43\plm0.10\km\kak119.2\plm0.6\km\kak5.621\plm0.026\kma\\
\km0.575\km\kak85.79\plm0.25\km\kak1276\plm5\km\kak42.27\plm0.16\kma&%
\km0.875\km\kak25.49\plm0.10\km\kak111.5\plm0.5\km\kak5.304\plm0.025\kma\\
\km0.585\km\kak88.66\plm0.25\km\kak1285\plm5\km\kak43.18\plm0.16\kma&%
\km0.885\km\kak25.49\plm0.10\km\kak104.9\plm0.5\km\kak5.038\plm0.023\kma\\
\km0.595\km\kak90.24\plm0.25\km\kak1282\plm5\km\kak43.61\plm0.16\kma&%
\km0.895\km\kak25.77\plm0.10\km\kak98.7\plm0.4\km\kak4.784\plm0.022\kma\\
\km0.605\km\kak91.38\plm0.25\km\kak1262\plm5\km\kak43.37\plm0.16\kma&%
\km0.905\km\kak26.20\plm0.10\km\kak93.1\plm0.4\km\kak4.550\plm0.020\kma\\
\km0.615\km\kak70.10\plm0.21\km\kak898.1\plm3.5\km\kak33.03\plm0.13\kma&%
\km0.915\km\kak26.81\plm0.10\km\kak87.6\plm0.4\km\kak4.322\plm0.019\kma\\
\km0.625\km\kak65.02\plm0.20\km\kak801.7\plm3.2\km\kak29.84\plm0.12\kma&%
\km0.925\km\kak27.49\plm0.10\km\kak82.8\plm0.4\km\kak4.117\plm0.018\kma\\
\km0.635\km\kak64.92\plm0.20\km\kak785.7\plm3.1\km\kak29.31\plm0.12\kma&%
\km0.935\km\kak28.57\plm0.10\km\kak78.74\plm0.33\km\kak3.950\plm0.017\kma\\
\km0.645\km\kak62.40\plm0.20\km\kak734.2\plm2.9\km\kak27.57\plm0.11\kma&%
\km0.945\km\kak29.86\plm0.10\km\kak74.74\plm0.31\km\kak3.780\plm0.016\kma\\
\hline
\end{tabular}}\vglue1mm
\caption{\label{tab:1}$\dd\sigma_{\pi\pi\gamma}/\dd M_{\pi\pi}^2$,
cross section and the pion form factor, in 0.01 $\GeV^2$ intervals.
The value given in the $M_{\pi\pi}^2$ column indicates the bin center.}
\end{table}
%
%

\vfill

\begin{thebibliography}{99}
\bibitem{vacpol} $\alpha(M_{\pi\pi})$ values are kindly provided by Fred Jegerlehner.
\end{thebibliography}
\end{document}